\documentclass{elsart}
\usepackage{graphicx,amssymb}
\usepackage[latin1]{inputenc}

\journal{Physica A}
\begin{document}

\begin{frontmatter}
\title{Deterministic walks in random networks: an application to thesaurus graphs}

\author{O. Kinouchi}
\author{A. S. Martinez}
\author{G. F. Lima}
\author{G. M. Louren\c{c}o}
\address{Departamento de F\'{\i}sica e Matem\'atica,  FFCLRP,
Universidade de S\~ao Paulo,  
Av. Bandeirantes 3900,  
CEP 14040-901, Ribeir\~ao Preto, SP, Brazil.}
\author{S. Risau-Gusman}
\address{Zentrum f\"ur Interdisziplin\"are Forschung, Wellenberg 1, D-33615
Bielefeld, Germany}

\begin{abstract}
In a landscape composed of $N$ randomly distributed sites in Euclidean 
space, a walker (``tourist'') goes to the nearest one that has not been 
visited in the last $\tau$ steps. 
This procedure leads to trajectories composed 
of a transient part and a final cyclic attractor of period $p$. 
The tourist walk presents a simple scaling with respect to $\tau$ and 
can be performed in a wide range of networks that can be viewed as 
ordinal neighborhood graphs.  
As an example, we show that graphs defined by thesaurus dictionaries share 
some of the statistical 
properties of low dimensional ($d=2$) Euclidean graphs and are easily 
distinguished from random link networks 
which correspond to the $d\rightarrow \infty$ limit.
This approach furnishes complementary information to the usual 
clustering coefficient and mean minimum separation length.
\end{abstract}
\end{frontmatter}

\maketitle

\section{Introduction}

We live in a world formed by networks: biological, social, linguistic and 
technological. 
The study and characterization of such networks have boosted recently
by the emergence of new ideas, increasing network databases and available
computational power to test models and link them to data \cite{strogatz}.

Networks form the substrate where dynamical processes can occur,
as the spreading of diseases or diffusion of information. 
Such processes are usually studied by using stochastic dynamics and 
random walks. 
Navigation processes and exploratory behavior \cite{viswanathan1,viswanathan2} 
can also be modeled by walks inside graphs. 
Of course, navigation processes are not
purely random nor purely deterministic. 
However, as a first step, it would be interesting to know the generic properties 
of deterministic navigation inside networks \cite{grassberger:92,gale:95,lima_prl2001}. 
If these networks are disordered, deterministic walks will lead to some statistics of 
trajectories which could provide information about the network topology and 
kind of disorder. 
Consider the following examples of deterministic walks in the
context of translation machines and thesaurus graphs.

Automatic translation software presents various problems due to the difficult 
task, even to humans, of converting phrases from one language to another.  
For example, it seems a desirable feature that if a translated sentence is 
translated anew to the original language one should get the original phrase. 
However, the iteration of the translation process in standard translator 
software frequently produces a drift from the original sense.  
After a transient, this drift 
achieves a two-cycle that constitutes a pair of usually nonsensical sentences, 
which are the same when translated back and forth. 

At the word level, a similar phenomenon occurs in analogical 
dictionaries or thesaurus.  
Starting from a random word, if one iterates the process going 
to the nearest synonymous in some sense (say, for example, the first word in 
the given list), one achieves also cycles of period two as shown in 
Table \ref{table:1}.  

\begin{table}[htb]
\begin{center}
\begin{tabular}{l}
\hline
1) {\bf link} $\rightarrow$ connection $\rightarrow$ {\bf link} \\
\hline
2) translation $\rightarrow$ conversion $\rightarrow$ {\bf change} $\rightarrow$ alter $\rightarrow$ {\bf change} \\
\hline
3) constitution $\rightarrow$  charter $\rightarrow$ contract $\rightarrow$ {\bf agreement} $\rightarrow$ 
accord  $\rightarrow$ {\bf agreement} \\
\hline
4) constitution $\rightarrow$  establishment $\rightarrow$ organization $\rightarrow$ 
association  $\rightarrow$ friendship $\rightarrow$ \\ 
companionship $\rightarrow$ company $\rightarrow$ 
corporation $\rightarrow$ business $\rightarrow$ commerce $\rightarrow$ trade $\rightarrow$ \\ 
deal $\rightarrow$ 
contract $\rightarrow$ {\bf agreement} $\rightarrow$ accord $\rightarrow$ {\bf agreement} \\
\hline
\end{tabular}
\end{center}
\caption{Examples of two-cycle obtained from iteration to the first word in synonymous
list (Microsoft Word 98): 1) zero transient trajectory; 2) trajectory with two transient 
steps; notice that trajectories may differ in 3) United Kingdom English and 4) USA 
English thesaurus. 
}
\label{table:1}
\end{table}

This iterative procedure converges, after some (sometimes large) transient time, 
to a two-cycle. Two-cycles are the universal attractors of this iteration process. 
This can be easily understood if one thinks about words in a thesaurus as nodes of a graph.  
This graph is not purely random but their nodes are semantically or 
etymologically linked defining some ordinal relationship in the node neighborhood (first 
neighbor, second neighbor etc.). Two-cycles appear when two mutually next 
neighbors are finally found. 

An interesting statistical question concerns the average degree of separation 
between two words in a thesaurus.  
After some experimentation in standard electronic thesaurus, one finds that the 
average degree of separation is of order $\log N$, where $N$ is the number of 
nodes (words) in the graph defined by a thesaurus.  
One also observes that synonymous have large clustering coefficient. 
This means that the probability that two synonymous of a word are also 
synonymous of each other is large. 
This suggests that a thesaurus is an instance of the so-called Small World 
networks \cite{watts}. 
This fact has been confirmed recently through exhaustive quantitative measures 
in the Merrian-Webster dictionary \cite{albert}, Roget's thesaurus \cite{steyvers} 
in the Wordnet Database \cite{steyvers,sigman} and even for free word associations 
database \cite{steyvers}. 

Here we suggest a different point of view to probe the structure of these highly 
non-trivial networks, through the study of the statistics of a simple 
deterministic walk on them: the ``tourist walk'' \cite{lima_prl2001}.  
Starting from an initial site, a walker (tourist) moves at each time step 
according to the simple deterministic rule: go to the nearest site that has not been 
visited in the preceeding $\tau$ time steps.   
The tourist performs a 
deterministic partially self-avoiding walk with memory window of $\tau$ steps.  
This iterative dynamics always leads to a trajectory that is composed of an initial 
transient part and a final attractor (a cycle of period $p \ge \tau + 2$) that 
traps the tourist.  
Here, $\tau$ is the limited tourist memory range 
or, alternatively, it can be thought as a refractory time of the sites.  
For $\tau = 0$, only cycles with period 2 appear, which correspond to a pair of mutually 
nearest neighbors points, corresponding to the two-cycle phenomena described earlier.  

The main result presented in this paper is that deterministic walks in spatial 
graphs (graphs whose nodes lie in Euclidean $d$-dimensional space) present some 
interesting statistical regularities. 
These regularities could serve as a benchmark when studying real world networks.  
In Section \ref{sec:deterministic_walks} we show that 
the tail of the distributions $P_{\tau}(p)$ of cycle period $p$, i.e., 
the number of different $p$-cycles divided by the total number of cycles, can be 
described in the whole $\tau$-range by a limited power-law.
In this way all curves $P_{\tau}(p)$ can be collapsed into a single universal curve 
independent of $\tau$. 
This is an extension and improvement to the description given in Ref. \cite{lima_prl2001}.  
Also, for $\tau = 0$, an analytical support, given by Cox's formula \cite{clark_1,clark_2,dacey,cox}, 
is presented along the text.
In addition to the quantities studied in Ref. \cite{lima_prl2001}, 
another quantity is examined: the probability $P_A(p)$  of a random site to 
belong to a $p$-cycle attractor basin. 
This probability is interesting since it shares the scaling properties of 
$P_{\tau}(p)$ and is easily measurable from experimental data. 
These quantities have been studied as a function of the dimensionality $d$ of 
the system where we show a slow convergence to the random link network. 
The random link network corresponds to the limit $d \rightarrow \infty$ and,  
for $\tau = 1$, the cycle distribution behaves as $P_{\tau}(p) \propto p^{-1}$. 
In Section \ref{thesaurus_graphs}, we compare our results, obtained in low dimensional Euclidean 
random graphs, to those obtained in the Small World graphs of thesaurus and find 
surprising similarities between them and huge differences from the $d \rightarrow \infty$ 
limit. 
This suggests that the short range connectivity structure of thesaurus graphs 
could be embedded in a low dimensional space, with the Small World behavior 
given by a small fraction of long range connections.
This contrasts to standand models which represent words in high dimensional 
Euclidean space \cite{steyvers,landauer:1997}.
Finally in Section \ref{conclusion} concluding remarks are presented. 

\section{Deterministic Walks}
\label{sec:deterministic_walks}

The construction of the $d$-dimensional Euclidean random graphs starts by randomly 
distributing the coordinates of $j = 1,2, \ldots, N$ sites with a uniform density $\rho$ in 
the interval $[0,1]$ in $d$ dimensions.  
The Euclidean matrix of distances $D_{ij}$
is used for ordering the neighbors and to create a 
neigborhood table $V_{ik} = j(k)$, where $j(k)$ is the site that is the 
$k^{\mbox{th}}$ nearest neighbor of site $i$. 

Notice that a random collection of points in Euclidean space
does not define a graph because a set of links is also necessary. 
If all the points are connected by links, we would have a totally connected 
graph. 
We will represent this graph by $G_{N-1}$, where $N-1$ is the number of outgoing 
links.
But with a memory window $\tau$, the walker dynamics is well defined if one 
knows what are the $\tau + 1$ nearest neighbors. 
Moves to the other neighbors never occur. 
This means that the walk is done inside a subgraph $G_{\tau+1}$ formed  by
linking directionally each site $i$ to its $\tau+1$ nearest neighbors. 
The tourist walk is always performed inside this directed subgraph that only 
presents an ordered set of neighbors $V_{ik}$ ($k=1,\ldots,\tau+1$) for each node. 
For example, in the $G_1$ subgraph each node is connected only to its nearest neighbor:
indeed this graph is composed by several disconnected parts, one for
each pair of mutual nearest neighbor.
In the $G_2$ subgraph, each node is linked to its two nearest neighbors. 
Such graph has a giant component (percolating cluster) containing more than $98\%$
of the nodes \cite{idiart:2001}.
Notice that $G_n$ is always a subgraph of $G_{n+1}.$

We stress that the Euclidean metric has been introduced in our model only as a 
simple mode of ranking the sites neighbors.  
As pointed out in Ref. \cite{lima_prl2001}, the dynamics is performed on 
the neighborhood table $V_{ik}$, not in the distance matrix $D_{ij}$.

\subsection{Walks without memory}

The density of cycles $D_{\tau}(p;d)$ is the number of different
$p$-cycles divided by $N$ for a given memory $\tau$ and dimension $d$.
As observed before, in the no memory situation $\tau = 0$, only $2$-cycles appear.
Nearest neighbor statistics in Poisson processes on 
Euclidean spaces are well known and have been used more intensively 
after the work of Clark and Evans \cite{clark_1}, followed by Refs.
\cite{clark_2,dacey,cox}.  
They have defined ``reflexive nearest neighbors'' as two sites that are 
the nearest neighbors of each other.  
These reflexive neighbors are precisely the attractors in the $\tau =0$ 
tourist walk.

Applying Cox's formula \cite{cox}, 
the density $D_{0}(2;d)$ of $2$-cycles 
in dimension $d$ is given by: 
\begin{eqnarray}
D_0(2,d) & = & \frac{P_{rn}(d)}{2} = \frac{1}{2(1 + p_d)} \; , \\
\label{eq:1}
p_d & = & 2 \; \frac{ \Gamma(d/2 + 1)}{\sqrt{\pi} \Gamma[(d + 1)/2]}
\; \int_{0}^{1/2} \mbox{d}x \, (1 - x^2)^{(d - 1)/2} \; , 
\end{eqnarray}
with $P_{rn}(d)$ being the fraction of reflexive neighbors and 
$\Gamma(z)$ the gamma function.
 
These formula can be worked out to obtain standard functions. 
Observing that:
\begin{equation}
\frac{ \Gamma(d/2 + 1)}{\sqrt{\pi} \Gamma[(d + 1)/2]} = \frac{1}{B[1/2,(d+1)/2]} \; ,
\end{equation}
where $B(a,b) = \Gamma(a) \Gamma(b) / \Gamma(a + b)$ is the beta function, 
one obtains:
\begin{equation}
   p_d  =   I_{1/4}(\frac{1}{2},\frac{d+1}{2}) \; ,
\end{equation}
with $I_z(a,b)$ being the incomplete beta function.
Some special values of these quantities are shown in Table \ref{table:2}.

\begin{table}[htb]
\begin{center}
\begin{tabular}{c|c|c}
\hline
$d$ & $p_d$ & $D_0(2,d)$\\
\hline
1 & 1/2  & 1/3 \\
2 & $(2 \pi + 3^{3/2})/(6 \pi)$ & $3 \pi /(2 \pi + 3^{3/2})$\\
3 & 11/16  & 8/27\\
\vdots & \vdots & \vdots \\
$\infty$ & 1  & 1/4 \\
\hline
\end{tabular}
\end{center}
\caption{Some values of $p_d$ and $D_0(2,d)$ as a function of $d$.}
\label{table:2}
\end{table}

\subsection{Walks in $d=2$}

We consider now the dependence on the memory $\tau$ in the two dimensional case.
For $\tau > 0$,  different period cycles coexist. 
Examples of cycles appearing for a landscape with $N = 400$ sites and several  
values of memory $\tau$ are given in Fig.~1, where open boundary 
condition has been used here. 
As $\tau$ grows, one observes the progressive disappearance, development or coalescence of 
cycles.

A natural question concerns the distribution of cycle periods 
\begin{equation}
P_{\tau}(p) = \frac{D_{\tau}(p)}{\sum_p D_{\tau}(p)} \; ,
\end{equation} 
and also if there is a value for $\tau$ where a percolation phenomenon occurs, i.e., 
the appearance of a cycle with diameter comparable to the system. 
However, the detailed study of this percolating regime demands 
strong computational effort, which is far from our resources. 
Here we only report results for the also interesting regime 
$\tau =  O(\log N)$, which is of the order of the mean minimum separation 
length in Small Word networks. 
This regime gives information about the clustering properties of the network
at scale of $O(\tau)$ cities.


Numerical simulations, using periodic boundary conditions, have been performed 
for $d=2$ and $\tau$ varying from $1$ to $10$ with $N=10^4$ and averaging over $100$
landscapes, see Fig.~2a. 
We have found that, for $p > 2 p_{min} = 2(\tau + 2)$, the probability  of 
appearance of a $p$-cycle $P_{\tau}(p)$ can be fitted by the expression:
\begin{equation}
P_{\tau}(p) = C(\tau) \: p^{- \alpha} \:\mbox{e}^{- \left[ p/p_0(\tau)\right]^2} \; , 
\label{eq:ptau}
\end{equation}
The fitting function gives $\alpha \sim 2.6 \pm 0.1$ independent of $\tau$. 
We observed that the cutoff lenght scales as $p_0 \propto \tau$, suggesting that
all these curves could be scaled into a single universal function 
which depends only on the ratio $p/\tau$. As the areas under the $P_\tau(p)$ curves
are always equal to one, we have found the data collapse on the function:
\begin{equation}
G(p/\tau) = \tau P_{\tau}(p) \; ,
\label{eq:sca_fun}
\end{equation}
as illustrated in Fig.~2b. 
It is worthwhile mentioning that other scalings as $(p - p_{min})/p_{min}$ or 
$p/p_{min}$ with $p_{min}= \tau+2$ 
have been tested and do not work as well as the one proposed above. 
A similar scaling does not hold for $d=1$ since in this case the exponent $\alpha$ 
depends strongly on $\tau$,  
see Ref. \cite{lima_prl2001}.


\subsection{High Dimensional Walks}

For $d>1$, the $P_{\tau}(p)$ curve slowly converges to the $d \rightarrow \infty$ 
behavior, which is also the behavior of the random link network 
(Fig.~3a),  which neglects the correlations among the distances between sites 
in a $d$-dimensional Euclidean space \cite{vannimenus,percus_thesis,percus}. 
The distribution tail has the form 
\begin{equation}
P_{\tau}(p) \propto p^{-\alpha(d)} \: \phi(p,\tau,N) \:,
\end{equation}
with $\alpha(d)$ varying from $\alpha(2)\approx 2.6$ to $\alpha(\infty) =1$ and $\phi(p,\tau,N)$
being a cutoff function.

\subsubsection{Convergence to the random link behavior}

The random link network has been generated numerically considering the 
distances between two points as uniform random variables in the interval $[0,1]$.
The distribution $P_{\tau}(p)$ for the random link network  
can be described by a power law $p^{-1}$ (see Fig.~3a) with a cutoff which
grows with the system size $N$, since it is a version of a random map model 
\cite{bastolla:1998}.
Notice, however, that we have studied a symmetrical situation 
($D_{ij} = D_{ji}$) with $\tau = 1$.
The Derrida random map \cite{derrida:2:1997} corresponds to an assymetrical random 
link case 
($D_{ij} \neq D_{ji}$), it presents cycle periods $p \ge 2$ even 
for $\tau = 0$ and the total number of attractors is of order $\log N$.
For $\tau = 0$, the symmetric case presents only two-cycles and is the 
correct approximation for the limit $d \rightarrow \infty$, with the number of attractors
scaling as $0.25 N$.


The slow convergence to the $d \rightarrow \infty$ behavior
also appears in other statistical 
quantities as, for example, the probability $P_A(p)$ that a random initial site 
leads to a cycle-$p$ (Fig.~3b).
When studying deterministic walks in real world networks, the quantity $P_A(p)$ 
is more convenient  from an experimental point of view. One can obtain better
statistics for $P_A(p)$ since each different initial condition gives valid data
and it is not necessary 
to keep track of only different cycles, as stated by the definition of 
$P_{\tau}(p)$. 

\section{Thesaurus graphs}
\label{thesaurus_graphs}

To exemplify the use of the tourist walk in ordinal ranking problems,
we have considered $\tau=0$ and $\tau = 1$ 
walks in a graph defined by a thesaurus. 
In this situation a word is chosen at random and uniformly from a standard 
dictionary.  
The chosen word is then placed in an electronic thesaurus (the USA English 
Microsoft 98 thesaurus).  
In the considered thesaurus, the synonyms are ranked according to their use frequency.  

\subsection{No memory}

For a $\tau=0$ walk, the more frequent used synonym of a given word (the first 
word in the synonymous list or ``first neighbor'') is then chosen as a new input and, 
again, its first neighbor is chosen in an iterative process, as 
illustrated in Table \ref{table:1}. 
This procedure leads to a transient and (generally) to 
a cycle of period 2, where two nearest reflexive synonyms are found. 
The distribution of transient times presents an exponential decay (Fig.~4). 

The proportion of reflexive neighbors could be  estimated in principle by 
measuring the 
fraction of initial words that are already in a two-cycle (like the word 
``link'' in Table \ref{table:1}). 
From 1000 initial words ($\sim 2\%$ of the whole dictionary), 
272 of them are found to belong to two-cycle.  
This value leads to $P_{rn} = 0.272 \pm 0.028$, with a confidence level $\gamma = 95\%$, 
which is very far from the range $[1/2,2/3]$ expected for random points in 
$d$-dimensional Euclidean space.  
However, it is clear from the distribution of the transient times 
(Fig.~4) that this number has been underestimated. 
One of the reasons for this underestimation is that a large fraction of two-cycles 
are composed by cycles containing two word expressions as: 
``{\bf punctual} - on time - {\bf punctual}''.  
Since we have started always from single word 
expressions, the cycle: ``{\bf on time} - punctual - {\bf  on time}'' will never 
be caught as a transient zero exemplar.
Another source for this discrepancy may be the presence of a small fraction of random links 
(see bellow). 
Extrapolating the exponential decay in transient to $t=0$ leads to a 
better estimate $P(t=0) = P_{rn} = 0.618$, 
which is compatible with a low dimensional $d = 2$ (or maybe $d = 3$) 
estimate by Cox's formula $P_{rn}(2) = 0.62$ ($P_{rn}(3) = 0.59$).

We shall also mention that some cycles of higher period have been found in the 
dictionary experiment: $(5.2 \pm 1.4) \%$ of three-cycles and 
$(0.5 \pm 0.4) \%$ of four-cycles with confidence level $\gamma = 95\%$. 
Thus, the examined dictionary may present links where $B$ is the 
nearest synonymous of $A$, $C$ is the nearest synonymous of $B$ and $A$ is the 
nearest synonymous of $C$. 
This kind of relation cannot be embedded in a metric space (due to the triangle 
inequality), suggesting that this thesaurus graph is slightly non-metric. 
We do not know if this is either a significant or desirable property, or an 
artifact from the specific dictionary examined.    


\subsection{One step memory}

For $\tau=1$ walks, we have found that the probability $P_A(p)$ of a 
word to belong to the basin of a $p$-cycle is very similar to that from walks 
in low dimensional Euclidean spaces, as shown in Fig.~5. 
This result is amazing and suggests that the lexicon graph could be embedded 
in a low dimensional Euclidean space preserving most of its 
neighbor order. 


Notice that we have worked only with the $G_1$ ($\tau=0$) and
$G_2$ ($\tau=1$) subgraphs, that is, graphs containing only one or two nearest neighbors of
each point. So, we are examining the local structure of the network, not the large
distance links which give the Small World behavior or power law degree of connectivity
\cite{steyvers,sigman}.
We have found that these subgraphs are very similar to those generated by a 
random collection of points in Euclidean space with $d = 2$ (or $d = 3$). It must
be emphasized that
higher values for $d$ are excluded because this would be reflected 
already in the $G_1$ and $G_2$ 
structure. The $d = 1$ case is also excluded once cycles with periods $5$, $6$ and $7$ have 
been measured and are forbiden by the model \cite{lima_prl2001}.

This result contrasts to standard models 
which represent words in high dimensional ($d > 100$)
Euclidean space, for instance, the Latent Semantic Analysis (LSA) 
model \cite{steyvers,landauer:1997}. If the present thesaurus were represented as points
in such high dimensional spaces, the tourist walk behavior should be similar to the
random link behavior. Instead,
our findings suggest that, for best representing
its local behavior, the nodes (words) from 
the studied thesaurus should be embedded in a low Euclidean dimension,
with a small fraction of assymetric links (to produce cycles with $p>2$ for $\tau = 0$).
Of course, as indicated by other studies \cite{steyvers,sigman}, there are also
long range links that give the Small World/Scale Free character of the network.

We have repeated this kind of study with the Portuguese Microsoft Word98
thesaurus. However, the existence of a large fraction of dead ends, i. e., 
nodes without outgoing links 
(reflecting a low quality database) prevented further analysis 
in this dictionary.

\section{Conclusion}
\label{conclusion}

We have shown that simple deterministic walks inside graphs with ordered 
neighborhood produce the emergence of cycles with interesting statistical 
properties.  
The spectrum of stable cycles $P_{\tau}(p)$ and the probability $P_A(p)$ of 
falling in a $p$-cycle from a random start give complementary information to 
measures like the average minimal length $L$ and clustering coefficient $C$. 
Networks with the same $L$ and $C$ could be distinguished by different 
spectrum $P_{\tau}(p)$ and $P_A(p)$.

Walks with no memory ($\tau =0$) easily detect absence of metrics in the graph. 
The curves $P_{\tau}(p)$ and $P_A(p)$ given here 
define class behaviors for the 
less informative spatial distribution of nodes (the Poisson case in Euclidean space) 
that can be used as a 
benchmark for results from walks in other networks. 
Since graphs with ranked neighborhood, but random structure, abound in 
the real world (from neural and ecological to social networks \cite{albert}), 
it could be interesting to find if deterministic 
walks on them belong, or not, to the same class defined by the Euclidean random 
graphs studied here. A surprising result is that the graph $G_2$ 
generated by a thesaurus is best 
represented by random points in low dimensional, even $d = 2$, Euclidean 
space.

\begin{ack}
The authors thank M. Idiart for sharing preliminary results with us, 
J. F. Fontanari for calling our attention to the random map model, Nestor Caticha,
A. C. Roque da Silva , N. A. Alves and  R. S. Gonzalez for fruitful discussions.  
O. Kinouchi acknowledges support from FAPESP.
\end{ack}

\bibliographystyle{elsart-num}
\bibliography{lima_phyA}

\newpage
{\bf Figure captions:}

{\bf Figure 1:} 
Examples of emergent cycles for $d = 2$ and $N = 400$. 
The memory values are: $\tau = 1, 3, 5$ and $9$. 
The same landscape has been used for each plot with open boundary condition. 
The growing, distabilization and fusion of cycles can be observed as a function of 
the memory $\tau$. Points not belonging to a cycle are not displayed.

{\bf Figure 2:}
a) Distribution of cycles periods $P_{\tau}(p)$ for $d = 2$ and different values 
of $\tau$ (from left to right, $\tau = 1, 2, 3, 4, 5, 6, 10$). 
Notice that $P_{\tau}(p)$ is defined only for integer $p$ 
(as in the $\tau = 1$ curve) nevertheless lines have been used for better 
visualization. 
b) Data collapse using Eq. \ref{eq:sca_fun}, $G(p/\tau) = \tau P_{\tau}(p)$ 
versus $p/\tau$ for 
$\tau = 1, 2, 3, 4, 5, 6, 10$ (from right to left).

{\bf Figure 3:}
a) Distribution of cycle periods $P_{\tau}(p)$ where the dotline represents 
the $P_{\tau}(p) \propto p^{-1}$ fit. 
b) Distribution of cycle periods $P_{A}(p)$.  
Curves are for $\tau = 1$ and from downwards to upwards with: 
$d = 2, 4, 8, 16$,  $32$ and symmetric random link. 
Other values are: $N = 10^3$ points averaged for $10^4$ landscapes 
with periodic boundary conditions.

{\bf Figure 4:}
Distribution of transient times for MS98 USA English thesaurus with 
$\tau = 0$ and $N=1000$ initial words. 
Fitting the points with $P(t) = c e^{-c t}$ leads to the extrapolation 
$P(0) = c = 0.618$ (open circle), which 
can be compared to $P_{rn}(2) = 0.6215$.

{\bf Figure 5:}
Comparison between fraction of sites $P_A(p)$ belonging to $p$-cycle 
attractor distribution of cycles for $\tau = 1$ in random Euclidean 
graphs with $d = 2$ (circles) for $N = 1000$ initial words of 
MS98 USA English thesaurus (triangles) and random link graphs (squares).

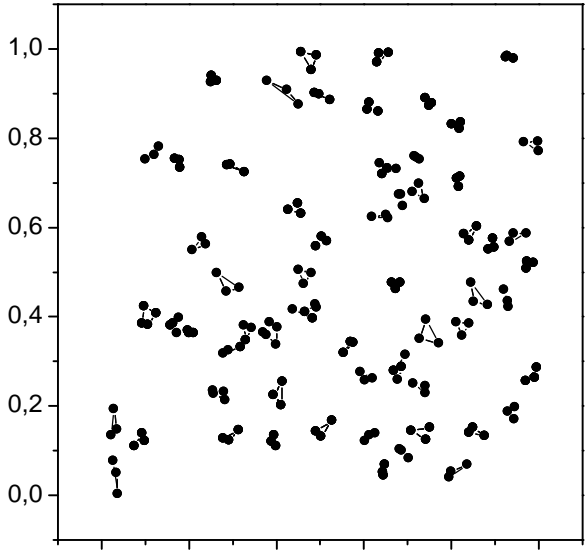
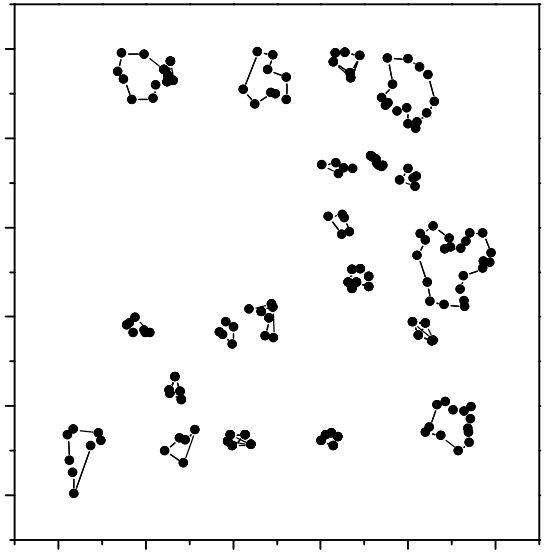
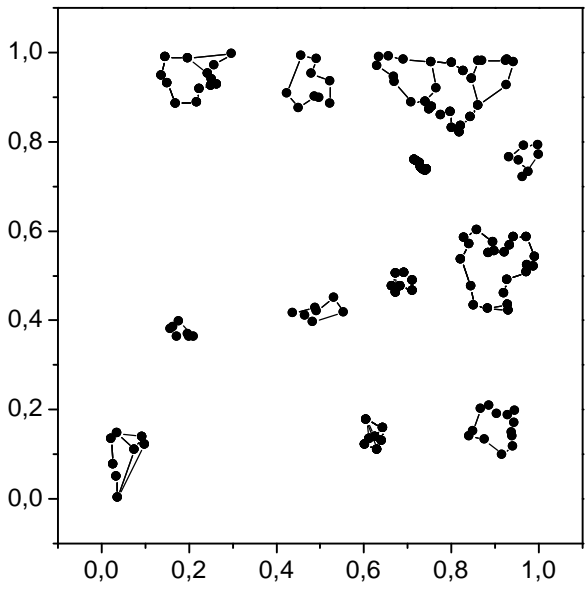
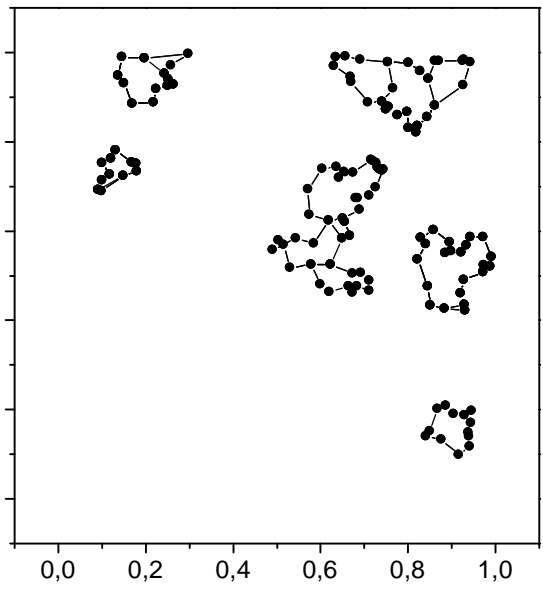
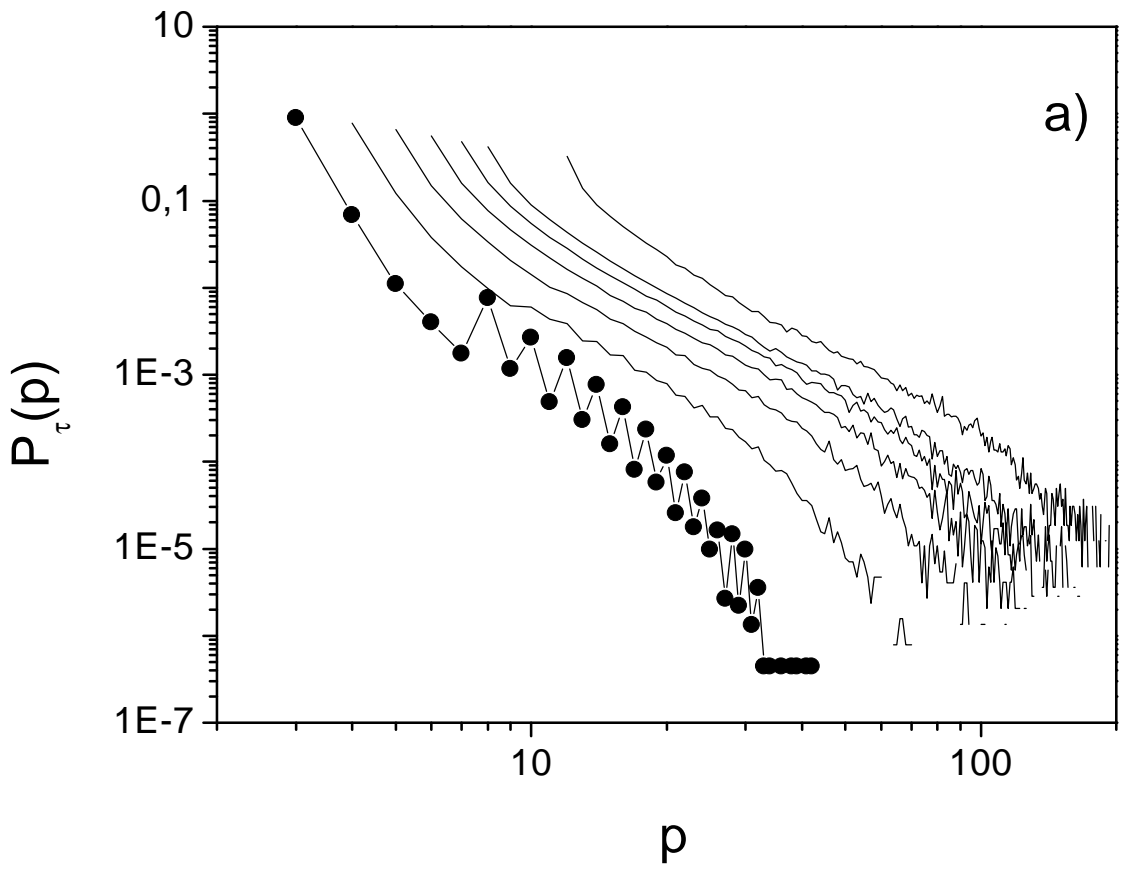
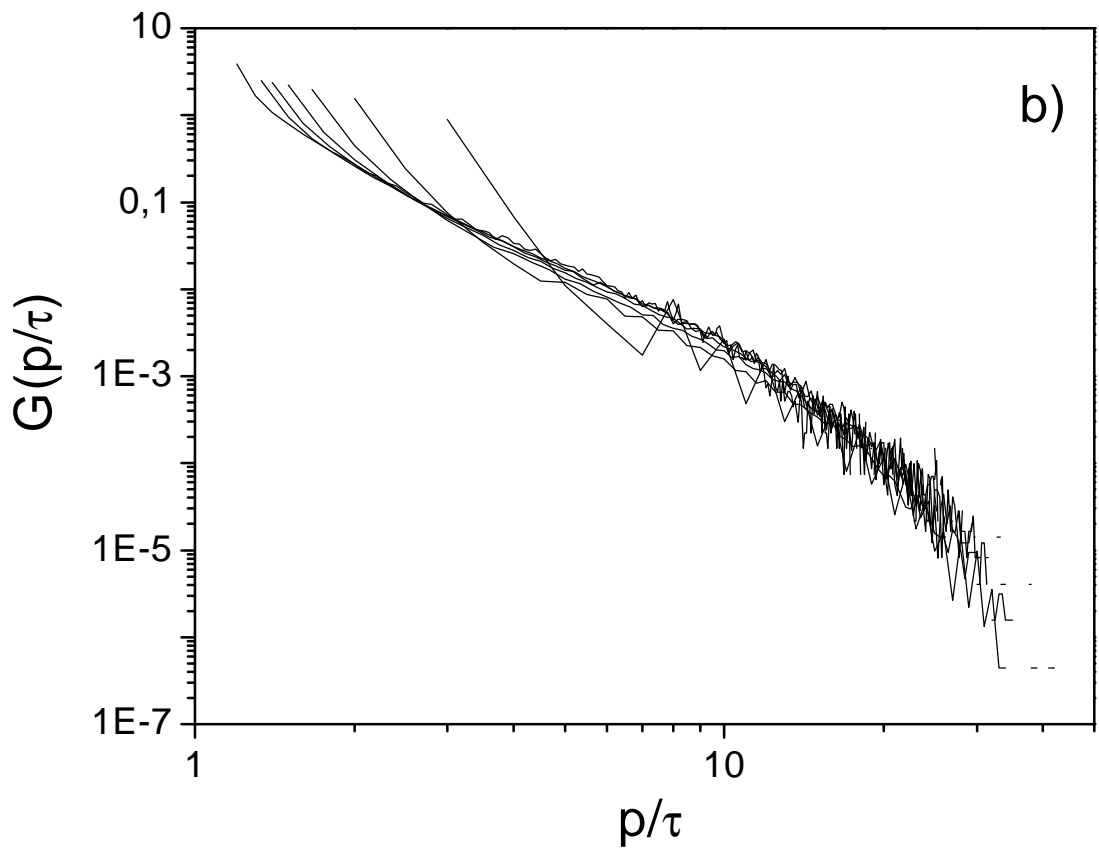
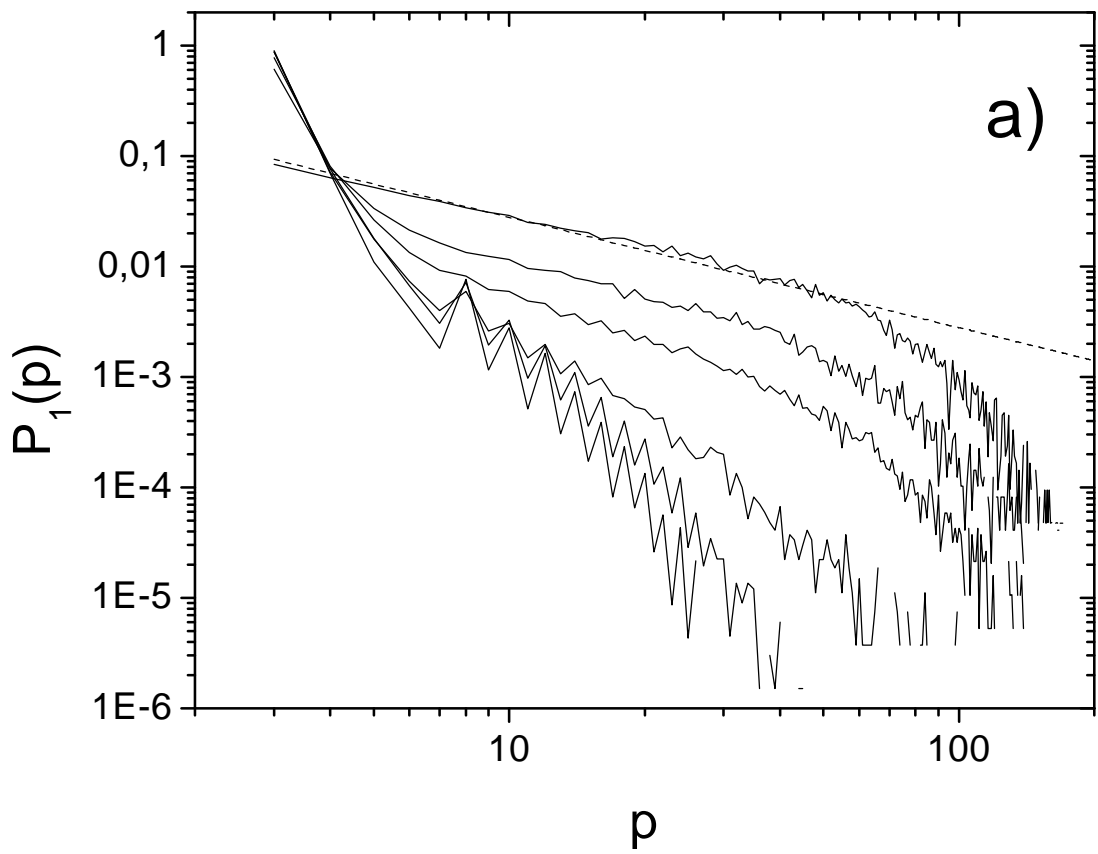
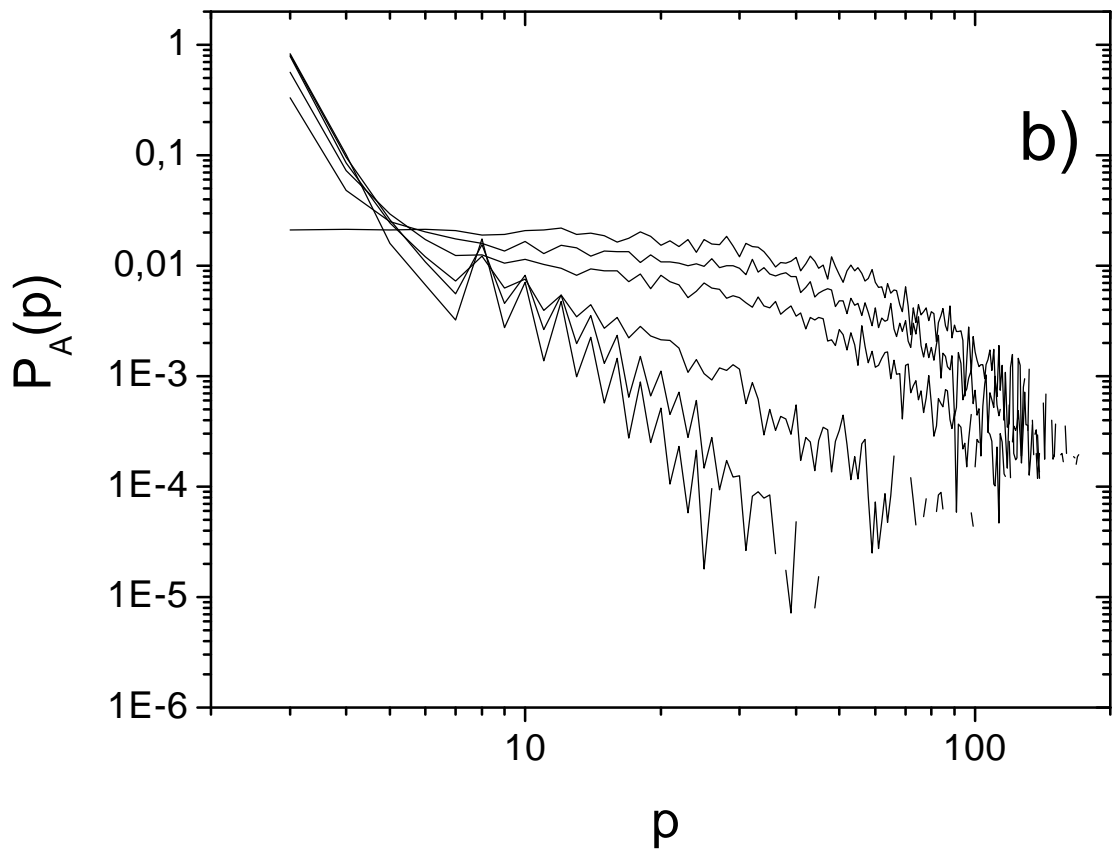
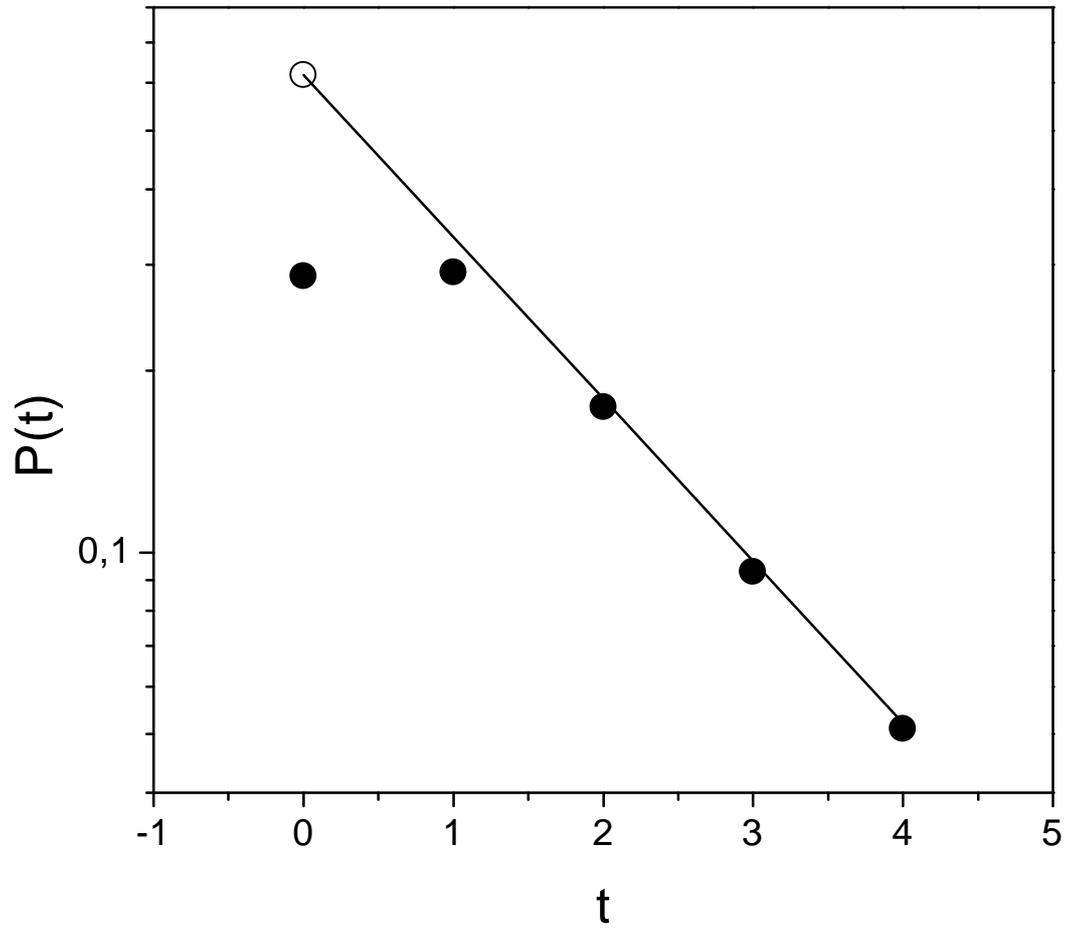
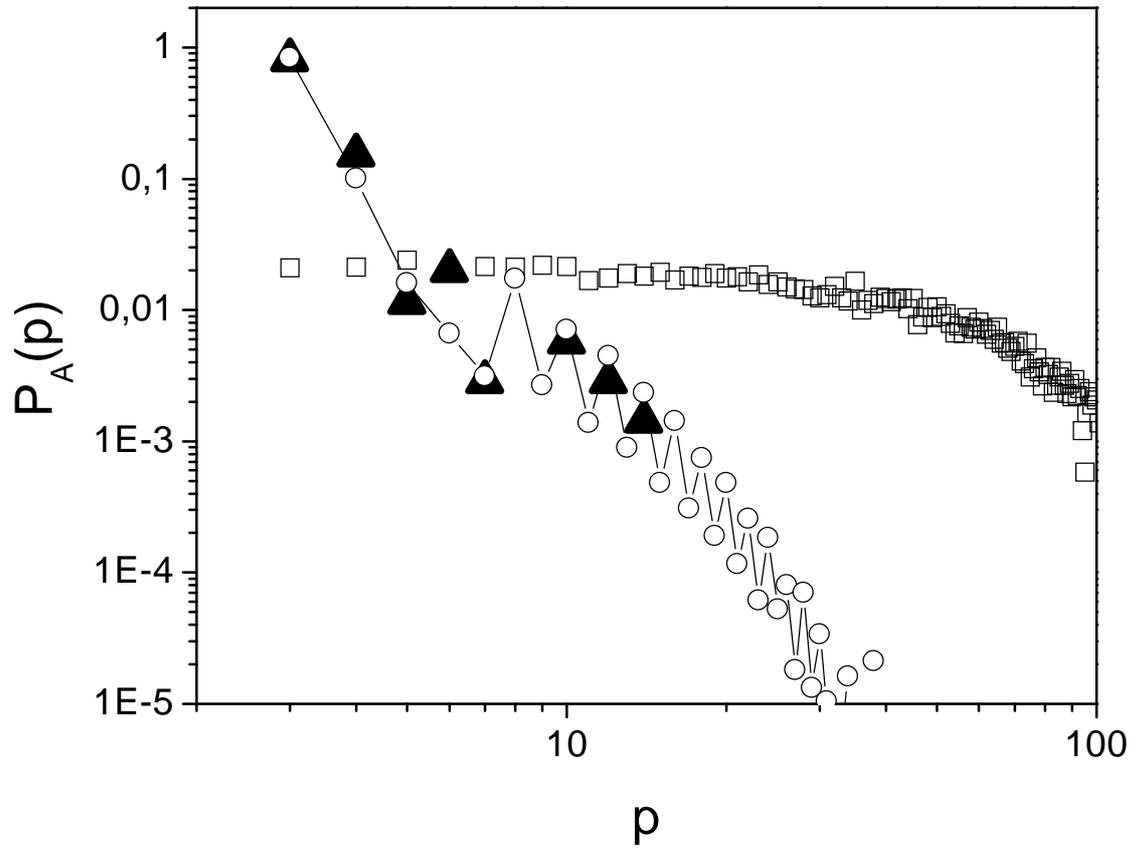
\end{document}